\documentclass[11pt,twoside]{article}

\usepackage{asp2014}

\aspSuppressVolSlug
\resetcounters

\begin{document}

\title{Moving Towards Greater Equity, Diversity, and Inclusion in Astronomy}

\author{Nuria~P.~F.~Lorente,$^1$ Mich\`ele~P\'eron,$^2$ and Jessica~Mink$^3$
\affil{$^1$Australian Astronomical Observatory, 105 Delhi Road, North Ryde, NSW 2113, Australia; \email{Nuria.Lorente@mq.edu.au}}
\affil{$^2$European Southern Observatory, Karl-Schwarzschild-Str. 2, D-85748, Garching bei M\"unchen, Germany}
\affil{$^3$Harvard-Smithsonian Center for Astrophysics, 60 Garden St. Cambridge, Massachusetts 02138, USA}}

\paperauthor{Nuria~P.~F.~Lorente}{Nuria.Lorente@mq.edu.au}{0000-0003-0450-4807}{Australian Astronomical Observatory}{}{North Ryde}{NSW}{2113}{Australia}
\paperauthor{Mich\`ele~Peron}{mperon@eso.org}{}{European Southern Observatory}{}{Garching bei M\"unchen}{}{D-85748}{Germany}
\paperauthor{Jessica~Mink}{jmink@cfa.harvard.edu}{0000-0003-3594-1823}{Harvard-Smithsonian Center for Astrophysics}{}{Cambridge}{Massachusetts}{02138}{USA}

\begin{abstract}
A diverse workforce and open culture are essential to satisfaction in the workplace, to innovation and creativity, and to the ability of an organisation to attract and retain talent. To ensure a diverse and inclusive workplace, efforts can be made to remove and prevent physical, systematic and attitudinal barriers. 

The invited talk on Diversity and Inclusion in Astronomy given by Mich\`ele P\'eron at this conference set the stage for discussion of this important topic, and presented how some of our institutions are addressing the issues and problems that exist, so as to set up a positive work environment for all.
The aim of this BoF was take some of the points raised by P\'eron and present them for discussion by the BoF participants. It was intended that the BoF be a forum for frank discussion and positive suggestions that participants could take back to their institutions.

\end{abstract}


\section{Introduction}

Equity, Diversity and Inclusion is an important topic which affects all of us, not just those from underrepresented groups. It impacts, positively or negatively, our community's technical and scientific creativity, performance, and productivity. 

Diversity of thought boosts innovation and efficiency and, we argue, mixed teams are more rewarding to work with. 
When tackling difficult challenges a diverse team performs better and more creatively than brilliant individuals~\citep{2007Page}. This is due to the perspective diversity (how situations are perceived) and heuristic diversity (how solutions to problems are generated) brought to the task. The different perspective/heuristic pair chosen by each member of the group leads to more potential solution paths than would be available to a homogeneous group of individuals, giving the diverse group the advantage. Further, Page argues that a non-diverse group of experts can be outperformed by a diverse group of people of moderate proficiency.

Inclusion is not about changing people in minority groups so that they will "fit in" to the existing culture, but about changing ourselves and our culture and environment to give everyone the support they need to succeed, irrespective of gender, race, disability, age, or any other metric. Inclusion is, at its heart, about truly respecting our colleagues, and using our differences in background and perspective to the benefit of all. An important step towards inclusion is developing an awareness of unconscious bias.

\section{Unconscious Bias}
Unconscious bias is a product of the brain's shortcut in the decision-making process. Stereotypes, mental constructs influenced by our culture and background, are used to very quickly assess people or to judge a situation, thus decreasing our mental load and freeing our attention for other tasks. Unfortunately the stereotypes we use are not always true or helpful, and can even be contrary to our explicitly held beliefs and values~\citep{1998Valian}. 
The result is that these biases can unconsciously guide our instincts and decisions away from fair, inclusive, and rigorous behaviours in situations when these are most required: during recruitment, staff appraisals, student grading, etc. \citep{2010Hill}.

Affinity bias, a specific example of unconscious bias, is particularly problematic when attempting to increase the diversity of teams through recruiting. This is the tendency most of us have to gravitate towards people who appear to be similar to us. In the recruitment context this can manifest as CVs which feel familiar: in education, career path, ethnicity, and gender. Consequently, it is easy to fall into the trap of hiring people who are like ourselves, thus propagating a lack of diversity within our teams, while at the same time wondering why we receive a lack of "suitable" applications from minority groups which we are consciously trying to attract.

Realising that we all have unconscious biases is the first step towards minimising their effect on our behaviour. Educating ourselves of what these biases may be (e.g.\ see the work of Mahzarin R. Banaji et al.~and the Harvard Implicit association test\footnote{\url{https://implicit.harvard.edu/implicit/}}) can increase our understanding our own biases and help us to compensate for them. 
Our belief in our own objectivity can be the greatest obstacle in facing up to and therefore working to minimise the effect of our (unavoidable) biases on our decisions~\citep[see][]{2014Urry}.

 \section{Snapshots of Diversity and Inclusion Work in the Astronomy Community}

The BoF began with the authors presenting the state of diversity and inclusion from the perspective of their institutions and countries. In the discussion that followed participants were invited to share the experience of their own institutions, pick up on points raised by the panel, and put forward ideas and suggestions of what we can do at an individual and institutional level, to improve diversity and inclusion to the benefit of all.

\subsection{The ESO Perspective}
ESO has a diversity and inclusion committee with a mission to formalise the many initiatives which have been implemented in the last decade and to develop an inclusion and diversity plan. It is important that this committee's membership mirrors the ESO population, rather than being composed primarily of people from minority groups, as diversity and inclusion is not an issue for and of minorities, but which affects everybody.

As in many organisations discussions on gender started amongst the astronomy groups much earlier than in the engineering sections. For some time special attention has been given to the composition of student and fellow selection boards. The female representation in these two groups oscillates between 30\% and 40\% year on year, with 28\% of ESO astronomers identifying as female, which is similar to that seen in other astronomy institutions. The fraction is much lower in the engineering groups (<10\%) and so a significant focus of the inclusion and diversity plan revolves around improving this number. 
We are also working on creating a more inclusive working environment, by introducing flexible working hours, the ability to work from home, and parental leave. 

The committee is also looking at the problem of vertical segregation. That is, the under-representation of women at leadership levels (currently 14\% of middle managers are women), and their relative over-representation in less senior positions. This vertical segregation cannot only be explained by horizontal segregation - that is the small number of women in some disciplines, and so work must continue on this.

\section{Strategies at the AAO}

The Australian Astronomical Observatory (AAO) has ongoing and evolving processes for assessing the performance of the organisation regarding equity and diversity, and addressing these through a range of initiatives.
A fundamental and essential requirement for this work has been the active support of the observatory's director, general manager, and executive team. From this, everything else flows, as it's very difficult to make substantial progress without it. This is because initiatives for real change often require funding or staff effort, and sometimes both.

To convert this intention to action, a Diversity Committee\footnote{\url{https://www.aao.gov.au/about-us/diversity-committee}}  has met regularly since 2011. It is this group's task to listen to staff, look at processes, find out what the problems are, and come up with solutions - ways in which we can foster positive change towards gender equity and diversity in our organisation. Importantly, it's not sufficient to put new strategies in place; we also have to be able to evaluate whether they are worthwhile, and change them if they are not.

An oft-repeated statement is that a barrier to increasing female representation in our engineering groups is that women simply don't apply for positions. 
Questioning this assumption, and investigating whether the recruitment policies and process may be contributors to the problem,
the AAO recruitment process was modified to include ensuring job advertisements are reviewed for biased language; requiring that the recruitment panel is itself diverse; providing the panel unconscious bias information, and encouraging them to take the Harvard implicit association tests to improve awareness of implicit bias, and to also be aware of their biases when looking applicant CVs and in the interview phase. 
Staff are asked to encourage individuals in underrepresented groups - particularly women to apply for open positions.
Finally, interview questions are decided in advance, to ensure that all candidates are given the same opportunity.

Flexible working arrangements, return to work scheme (for staff who take extended leave for caring responsibilities), and training in 
unconscious bias and diversity have also been successful initiatives, which now form part of the workplace culture.

\section{Diversity Work in the AAS}

The American Astronomical Society (AAS) has several committees tasked with advising the society on the subject of diversity and inclusion. These are the Committee on the Status of Women in Astronomy (CSWA), the Committee on the Status of Minorities in Astronomy (CSMA), the Sexual orientation and Gender identity Minorities in Astronomy (SGMA), and the Working Group on Disablities in Astronomy (WGDA).

In 2015 an Inclusive Astronomy Conference was held to discuss the issues faced by underrepresented groups in the US astronomy community. This resulted in a set of recommendations (known as the "Nashville Recommendations"\footnote{AAS Nashville Recommendations: \url{http://bit.ly/2r3nidq}}) being made to the Society
in the subjects of barriers to access, inclusive climates, policy and leadership, and inclusive practice, spanning the short, medium, and long terms. These recommendations were endorsed in 2016, and an aim to "Identify short-term, medium-term and long-term goals based on recommendations relevant to the institution and people at the institution, develop and commit to individual, group, and institutional plans" was adopted.

Work on improving diversity and inclusion continues, thanks to the efforts of many people within the community. A recent highlight is the release of the Safe Meeting Policy, in March 2017, at the recommendation of the SGMA. This aims to ensure the safe participation of people identifying as LGBTQIA in all AAS meetings, by
requiring meetings to be located in cities where LGBTQIA attendees would be welcomed and accommodated.

\section{Conclusion}
This BoF was very well attended, and we thank all those who were involved for their active and frank participation in the discussion. Of course the conversation does not end with the BoF, and we look forward to future BoFs and other ADASS sessions where these issues can be discussed and progress in the area made, for the benefit of the entire community.

\bibliography{B2}  

\end{document}